\PassOptionsToPackage{breaklinks}{hsperref}
\documentclass[10pt, conference, letterpaper]{IEEEtran}

\usepackage[T1]{fontenc} 
\usepackage[utf8]{inputenc}
\usepackage{cite}
\usepackage{amsmath,amssymb,amsfonts}
\usepackage{algorithmic}
\usepackage{graphicx}
\usepackage{textcomp}
\usepackage{balance}
\usepackage{microtype}
\usepackage{subcaption}
\usepackage{units}
\usepackage{enumitem}
\usepackage{xurl}
\usepackage{tikz}
\usetikzlibrary{arrows,shadows,positioning,backgrounds}
\usepackage{booktabs}
\usepackage{arydshln}
\usepackage{makecell}
\usepackage{dblfloatfix}

\newcommand*\circled[1]{\tikz[baseline=(char.base)]{
            \node[shape=circle,draw,inner sep=1.5pt] (char) {#1};}}

\newcommand{\ie}{i.e.,}
\newcommand{\eg}{e.g.,}
\makeatletter
\newcommand{\etal}{et~al\@ifnextchar.{}{.\@}}
\makeatother

\newcommand{\afblock}[1]{\noindent{\textbf{#1}}} 
\newcommand{\takeaway}[1]{\noindent{\textbf{Takeaway.}} \textit{#1}}
\newcommand{\sref}[1]{Section~\ref{#1}}
\newcommand{\fig}[1]{Figure~\ref{#1}}

\usepackage{xcolor}
\makeatletter

\newcommand\thefontsize[1]{{#1 The current font size is: \f@size pt\par}}
\makeatother

\newcommand\copyrighttext{%
  \footnotesize © IFIP, (2019). This is the author's version of the work. It is posted here by permission of IFIP for your personal use. Not for redistribution. The definitive version was published in \emph{Network Traffic Measurement and Analysis Conference (TMA), 2019}.}
\newcommand\copyrightnotice{%
\begin{tikzpicture}[remember picture,overlay]
\node[anchor=south,yshift=10pt] at (current page.south) {\fbox{\parbox{\dimexpr\textwidth-\fboxsep-\fboxrule\relax}{\copyrighttext}}};
\end{tikzpicture}%
}

\begin{document}
\bstctlcite{IEEEexample:BSTcontrol}

\title{An Empirical View on Content Provider Fairness}

\author{\IEEEauthorblockN{Jan R\"uth, Ike Kunze, Oliver Hohlfeld}
\IEEEauthorblockA{\textit{RWTH Aachen University} \\
\textit{\{rueth, kunze, hohlfeld\}@comsys.rwth-aachen.de}
}
}

\maketitle

\begin{abstract}
Congestion control is an indispensable component of transport protocols to prevent congestion collapse.
As such, it distributes the available bandwidth among all competing flows, ideally in a fair manner.
However, there exists a constantly evolving set of congestion control algorithms, each addressing different performance needs and providing the potential for custom parametrizations.
In particular, content providers such as CDNs are known to tune TCP stacks for performance gains.
In this paper, we thus empirically investigate if current Internet traffic generated by content providers still adheres to the conventional understanding of fairness.
Our study compares fairness properties of testbed hosts to actual traffic of six major content providers subject to different bandwidths, RTTs, queue sizes, and queueing disciplines in a home-user setting.
We find that some employed congestion control algorithms lead to significantly asymmetric bandwidth shares, however, AQMs such as FQ\_CoDel are able to alleviate such unfairness.
\end{abstract}

\section{Introduction}
\copyrightnotice
The Internet has grown way beyond its original purpose of being a research network.
Today, thousands of autonomous systems connect and exchange data.
The fundamental principles governing this data exchange are well established since decades and defined in IETF RFCs.
To this end, the current best-effort Internet relies on CC to \emph{i)} not collapse the network, and to \emph{ii)} achieve fairness for flows competing for bandwidth at a bottleneck.
For TCP, RFC 5681~\cite{RFC5681} requires the implementation of slow start, congestion avoidance, fast retransmit, and fast recovery (generally known as TCP Reno).
Other algorithms improve on certain aspects of Reno, \eg{} to enable higher performance over large BDP networks.
Usually, when a new or modified CC algorithm is proposed, it is rated in terms of TCP fairness when competing with Reno or Cubic as Linux's default CC algorithm.
While fairness is generally a hard to define property for Internet flows and flow-rate fairness is a controversial metric~\cite{brisco07:ccr}, it is still widely used.
In 2005, Medina~\etal{}~\cite{medina05:ccr} showed that \emph{most} Internet flows halve their congestion window on loss and are thus TCP conform, leading to an expected flow-rate fairness~\cite{RFC5290}.

Since then, the Internet landscape has drastically changed, end-users use the Internet with increasing access speeds~\cite{akamai_stateoftheinet} and content such as videos is causing a substantial fraction of Internet traffic~\cite{trevisan18:conext:ispedge,erman11:imc:videocellular}.
These increasing demands have led to a logical centralization of the content-serving Internet where a few big players serve the majority of the content~\cite{carisimo17:tma:inetcore,labovitz10:sigcomm:interdomain}.
In previous work~\cite{rueth18:tma}, we have shown that CDNs specialize in serving such content by tuning their TCP stacks beyond RFC-recommended values in hope for higher performance and user satisfaction.
Fundamentally, such observations raise the question of fairness, and in fact, from an economic standpoint being unfair to a competing CP might be advantageous (\eg{} by being able to deliver data with more than a fair bandwidth share).
While identifying a CP's CC algorithm (\eg{} via~\cite{yang11:icdcs:ccident}) helps in understanding its principal behavior, these works do not take into account the actual parameterization of the algorithms which have the potential of drastically changing the fairness.
Transport protocol evolution with QUIC has the potential to further lower the hurdle for modification in the future, given its realization in userspace for flexible customization.

In light of these historical changes, this paper investigates the behavior and interaction of large CPs. %
Thereby, we shine a light on current practices and evaluate the question of whether actual Internet traffic adheres to the conventional understanding of fairness.
To this end, we devise a methodology that enables us to compare testbed results with actual Internet traffic.
Specifically, this work contributes:
\begin{itemize}[noitemsep,topsep=0em,leftmargin=10pt]
	\item We present a testbed methodology using RTT-fairness to study actual TCP traffic by major CPs to account for a broad set of TCP optimizations used in practice.

	\item We compare fairness properties of testbed hosts to actual traffic by six major CPs subject to different bandwidth, RTT, queue sizes, and queueing disciplines in a home-user setting.
	We find that achieving a fair bandwidth share largely depends on the competing congestion control algorithms (Cubic vs. BBR) and the involved network conditions.
\end{itemize}
\afblock{Structure.}
We introduce flow-rate fairness and related works in Section~\ref{sec:bg}.
We then introduce our testbed methodology and its validation in Section~\ref{sec:testbed}.
Section~\ref{sec:results} discusses the results of our fairness study before we conclude the paper.

\section{Background and Related Work}

\label{sec:bg}
One of the key challenges in the Internet is the decentralized resource allocation of bandwidth.
However, TCP's initial design only prevented overloading single end-points and did not consider the possibility that the network itself could become overloaded and collapse upon this congestion.
As centralized algorithms are not deployable on the Internet, decentralized CC was soon added to TCP's design.
However, the highly distributed nature of the Internet quickly showed that there are scenarios where the early CC often yields less than optimal performance which has led to a plethora of research for evolved and optimized algorithms.
With the introduction of ever more algorithms, questions about their interaction arose challenging how these algorithms share the available bandwidth.
Research has hence also considered these aspects by investigating \emph{fairness} of CC.
Well-studied fairness measures are the intra-protocol flow-rate fairness, \ie{} how well do two instances of the same algorithm share the available bandwidth, the RTT-fairness, \ie{} what happens if the flows have different RTTs, and the inter-protocol fairness, where two instances of two different algorithms are investigated.

\afblock{Intra-Protocol and RTT-Fairness.}
For Cubic, research has commonly found decent intra-protocol fairness and an inverse-proportional RTT-fairness, meaning that instances with smaller RTTs get a larger share of the overall bandwidth.
These findings have been confirmed for a large set of different network characteristics, ranging from small (\unit[10]{Mbps}, \eg{}~\cite{leith08:cubicexperiment}) to large bottleneck bandwidths (\unit[10]{Gbps}, \eg{}~\cite{xue14:heterogeneousTCP}) or short (\unit[16]{ms}, \eg{}~\cite{ha08:cubic,miras08:highSpeedFairness}) to long (\unit[324]{ms}, \eg{}~\cite{ha08:cubic,miras08:highSpeedFairness}) RTTs.

For BBR, less research exists and the available studies partly disagree on the properties of BBR.
This is especially true for intra-protocol fairness, as Cardwell~\etal~\cite{cardwell16:ietf97iccrg:bbr} claim a high degree of fairness across the board, while Hock~\etal~\cite{hock17:icnp:bbrEval} identify scenarios where the fairness is significantly impaired.
Regarding RTT-fairness, it is commonly found that BBR has a proportional RTT-fairness property, \ie{} a flow with a larger RTT gets a larger share of the available bandwidth~\cite{cardwell16:ietf97iccrg:bbr,ma17:congestionBasedFairness,scholz18:bbrDeeperUnderstanding}.
Hock~\etal~\cite{hock17:icnp:bbrEval} generally confirm the findings but by investigating two different bottleneck queue sizes, they find that in scenarios with a smaller queue size (0.8$\times$ bandwidth delay product (BDP)) flows with a smaller RTT have a slight advantage, while in large buffer scenarios (8$\times$BDP) the inverse is true and flows with larger RTTs have a significant advantage.

\afblock{Inter-Protocol Fairness.}
While the intra-protocol and RTT-fairness of CC is important for a large scale-out of the algorithms, the inter-protocol fairness property shines a light on the coexisting use of different CC algorithms in the Internet.
Unfortunately, several groups of researchers have found that BBR and Cubic do not cooperate well, as Cubic flows dominate BBR flows in scenarios with larger buffers (generally above 1$\times$BDP) while the opposite is true for small buffer scenarios~\cite{cardwell16:ietf97iccrg:bbr,hock17:icnp:bbrEval,scholz18:bbrDeeperUnderstanding}.

While many studies investigate how certain algorithms affect each other, there is missing up to date research on which are actually used in the Internet.
Moreover, many studies neglect the parameterization and tuning potential of these algorithms that are used in practice.
To address this, this study explores if actual Internet traffic of large content providers---which carry the bulk of today's Internet traffic---still adheres to the conventional understanding of TCP fairness.

\section{Methodology}
\label{sec:testbed}

CC research traditionally involves {\em simulation} or {\em testbed} studies, which give researchers {\em complete control} over the investigated scenarios.
While this is desirable for controlled experiments, the involved abstractions and assumptions do not allow to completely cover real-world settings.
For example, the employed algorithms and their parameterization in real-world systems are typically unknown.
To study CC fairness in practice, we, therefore, contact real-world Internet systems with a testbed setup.
This enables us to still control {\em some} parameters (\eg{} bottleneck bandwidth and delay) while studying the CC algorithms as run by real systems.
This way we can study if Internet traffic by CPs still adheres to the conventional understanding of TCP flow-rate fairness.

\subsection{Home User (Residential Access) Scenarios}
\begin{figure}[t]
\center
\includegraphics[]{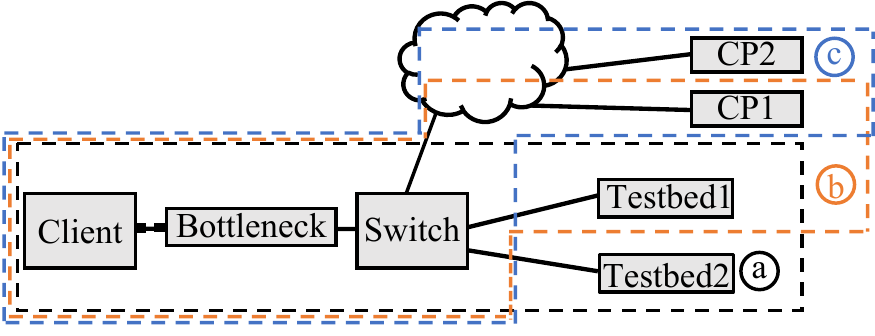}
\caption{Testbed topology with testbed and online components. Scenario\,\protect\circled{a} (testbed-only), Scenario\,\protect\circled{b} (testbed \& Internet), Scenario\,\protect\circled{c} (Internet-only).}
\label{fig:testbed_overview_complete}
\vspace{-1em}
\end{figure}

The fundamental design choice of our study is to investigate fairness from the perspective of an end-user accessing the Internet through a home router.
Even though peering links have been identified as possible points of congestion~\cite{dhamdhere18:sigcomm:Interdomcongestion}, it is still widely believed that access links form the bottlenecks and thus congestion happens at the network edges, more specifically at the end-user's access link~\cite{bauer09:tprc:ccevolution}.
We model this scenario in the form of a simple dumbbell topology that is the foundation of our testbed which we illustrate in \fig{fig:testbed_overview_complete}.
The user---represented by the client---is connected to the testbed network via a dedicated machine serving as a configurable bottleneck (via Linux's traffic control (TC) subsystem).
In general, the client can request traffic from all kinds of sources, from within the testbed and from Internet sources.
For our study, we focus on three distinct scenarios.

In Scenario\,\circled{a} (testbed only), we investigate the out-of-box performance of CC by simultaneously requesting traffic from two testbed machines.
This, above all, establishes a baseline and identifies potential influencing factors on the overall interaction of CC.
Building upon this baseline, Scenario\,\circled{b} (testbed \& Internet) then replaces one of the two testbed flows with a flow originating from the Internet.
Thus, we compare how the Internet flows interact with the out-of-box CC algorithms.
Finally, Scenario\,\circled{c} (Internet-only) considers the case where both flows originate from the Internet to investigate how and whether their interactions differ from the previous scenarios.

The common goal of all three scenarios is to judge the bandwidth sharing behavior of different CC algorithms in different network settings for which we consider four network characteristics.
The bottleneck \emph{bandwidth} and the overall \emph{RTT} hereby give hard upper bounds (in terms of available data rate) and lower bounds (in terms of responsiveness) on the overall performance, while the bottleneck queue, characterized by its \emph{queue size} and \emph{queuing discipline}, introduces jitter, and loss.

\subsection{Testbed Setup} %
\label{sub:architecture}
The core of our testbed consists of one machine that represents the end-user and hence serves as the client throughout the scenarios and another machine which represents the user's access link and hence the bottleneck of the overall connections.
The latter is then used to model all connection-specific properties like delay or bandwidth.
For the scenarios where we create flows from within our network, we deploy one machine for each flow that is involved and configure server-side parameters like the deployed CC algorithm on them. 
All machines within the testbed use a Linux 4.13 kernel and they are interconnected via Gigabit Ethernet to ensure that the physical links never become a bottleneck.

\afblock{Limiting Bandwidth.}
Most configurations, like rate-limiting, are done on the bottleneck's egress queues.
Here, we configure the bandwidth and queue size using a token bucket filter with a burst size of a single frame while using different queue management techniques.
Even though Internet access links are often asymmetrical, we disregard this fact as we are not interested in investigating reverse-path congestion and use the same bandwidth in both directions.

\afblock{Ensuring RTT-Fairness.}
To reason about the main question of this work, \ie{} about the bandwidth sharing properties of Internet flows, we employ the RTT-fairness property which states that two flows should share the bandwidth equally if they have the same RTT.
This, in turn, means that we only consider those cases where the different flows have the same RTT and we consequently use fairness synonymously for RTT-fairness.

To add delay to our testbed, we use TC to perform ingress packet processing at our bottleneck.
There, we redirect traffic to an intermediate queue disc enabling us to use NetEm to add a delay before we release the packet for forwarding to the actual egress queue.
We do not configure any artificial jitter using NetEm as this causes packet reordering; the additional delay and jitter are thus only caused by the egress queue.
To have symmetric delays, we add half of the configured delay to each ingress of the bottleneck.
While care needs to be taken in sizing the NetEm queue to not cause artificial packet loss, this approach has the advantage that the end-host stacks are not involved in the delay which is known to badly interfere with CC when Linux detects queuing pressure (TCP small queues)~\cite{google_qdisc}.
Further, in Scenarios\,\circled{b} and \circled{c} we even have no control over all end-hosts.
To ensure that we can investigate RTT-fairness, we set different delays for each flow to harmonize their RTTs.
To this end, we measure the \emph{minimum} RTT through our testbed (using TCP pings) when not using any artificial delays for each flow.
We then use each flow's min RTT to configure delays such that all flows experience the same artificial min RTT.

\afblock{Limitations.}
Our testbed has several limitations that need consideration.
We must ensure that our traffic shaper is the actual bottleneck of the path from the CP to our client.
Since we do not have full control over all involved entities, we can only configure bandwidths that are sane given our interconnection.
Our testbed uses Gigabit Ethernet, our Institute is then connected via \unit[10]{Gbps} to our University's backbone, which in turn connects to the German research network (DFN) via \unit[40]{Gbps} which then peers at DE-CIX with all CPs investigated in this study.
Thus, shaping traffic for typical end-user access links should render the bottleneck to our traffic shaper.
Further, we need to artificially bump up the RTTs at least to the largest minimum RTT measured.
For us, the CPs typically show RTTs around \unit[5]{ms} to \unit[10]{ms} which enables us to investigate a large range of RTTs.

Additionally, to ensure repeatability and independence, we take several precautions to avoid undesired side-effects. 
First, to investigate the interaction of CC, we must be actually limited by the congestion window which is why we advertise an initial flow-control receive window of 200 segments.
In the same way, we ensure that send and receive buffers are large enough to fully utilize the available bandwidth and do not introduce an undesired new bottleneck.
Finally, we clear all TCP caches after each measurement to ensure that cached metrics such as ssthresh do not affect future measurements (testbed only).

\begin{table}
	\centering
	\begin{tabular}{@{}l|l}
		\toprule
		Setting & Parameter Space\\
		\midrule
		Bandwidth & \unit[50]{Mbps}, \unit[10]{Mbps}\\
		RTT & \unit[50]{ms}, \unit[100]{ms}\\
		Buffer sizes & $0.5\times\text{BDP}$, $2\times\text{BDP}$\\
		Queueing discipline & drop-tail, CoDel, FQ\_CoDel\\
		\bottomrule
	\end{tabular}
	\caption{Study parameter space}
	\label{tab:parameterspace}
	\vspace{-2em}
\end{table}

\subsection{Parameter Space}
Selecting reasonable parameters for our testbed is challenging.
We must adhere to the testbed's limitations while seeking to replicate a reasonable end-user environment.
Table~\ref{tab:parameterspace} summarizes the parameter space which we discuss next.

\afblock{Bandwidth.}
To ensure that the bottleneck link is within our testbed, we have to set the bottleneck link bandwidth accordingly.
To identify the bandwidth provided by the individual CPs, we performed a larger number of bandwidth tests to determine which data rates are reliably offered by the different CPs.
We have found the lowest data rates to be around 60 Mbps.
Adding a safety margin, we choose 50 Mbps as our upper data rate limit which according to Akamai~\cite{akamai_stateoftheinet} is representative for mid-sized access links.
Further, we choose 10Mbps as a lower bound to represent a low-end connection.

\afblock{RTTs.}
We choose \unit[50]{ms} as the lower bound for the minimum RTT and \unit[100]{ms} as a representative for higher latencies, even though we expect typical CPs to usually have much lower RTTs to their customers.
However, these increased RTTs make it possible to reduce the relative error when we pad up the RTTs to ensure RTT-fairness between connections.

\afblock{Buffer Sizes.}
For the bottleneck buffer, we experiment with different queue sizes since we know of no study that investigates typical last-mile buffer sizes.
While the potential for overly large buffers (bufferbloat) is known~\cite{gettys12:bufferbloat}, less than 1\% of the end-user flows were observed to experience RTT variations larger than \unit[1]{sec} by a major CDN~\cite{hohlfeld14:imc:bufferQoE}.
Therefore, we choose one overly large buffer in the order of 2$\times$BDP and, inspired by research advocating new buffer sizing rules ($\sqrt{num\_flows}$~\cite{appenzeller04:sigcomm:SqrtN} and $\log{win\_size}$~\cite{enachescu06:infocomm:LogW}), one smaller buffer size of 0.5$\times$BDP which, for our investigated bandwidths and delays, yields queue sizes between both proposed sizing rules.

\afblock{AQM.}
In addition to these parameters, we also change the queuing discipline between a regular drop-tail queue and (FQ-) CoDEL~\cite{nichols12:queue:codel} to investigate the impact of AQM on fairness.

\subsection{Fairness Metric}
We rate the fairness by capturing the traffic that the client receives for each flow.
To this end, the client requests a first flow from one machine and after \unit[5]{s} a second flow from another machine.
Both flows then continue to transmit data for another \unit[40]{s} before shutting down.
Of the overall \unit[45]{s}, we investigate \unit[35]{s} starting \unit[5]{s} after the second flow starts its transmission.
While this methodology is above all intended to focus on the long term fairness between the two flows, we also examine whether it is important which flow is started first by including experiments with a flipped starting order.
We repeat each measurement 30 times to investigate the stability of our results.

To rate the fairness between both flows, we look at the ratio of transmitted bytes (over the same timespan of the shorter flow) and define our fairness measure as:
$$
fness(a,b) = 
     \begin{cases}
       \phantom{-}1 - \frac{bytes(a)}{bytes(b)} &\quad\text{if bytes(b)} \ge \text{bytes(a)}\\[0.5em]
       - 1 + \frac{bytes(b)}{bytes(a)} &\quad\text{if bytes(a)} > \text{bytes(b)}\\ 
     \end{cases}
$$

Intuitively, \emph{fness}(A,B) maps the fairness behavior of the two flows into the range of [-1, 1] with zero indicating absolute fairness, -1 that Flow\,A absolutely dominates Flow\,B and a value of 1 the opposite.
In between, the measure depicts the ratio of bytes actually transmitted, \eg{} 0.5 indicates that Flow\,B transmitted twice the bytes compared to Flow\,A.

\subsection{Testbed Validation}
To investigate if our testbed produces meaningful results, we seek to confirm known findings about the behavior of CC with our testbed. 
Due to the fact that related work considers a wide range of parameter settings and different variations of dumbbell topologies, we do not aim to exactly replicate specific results of related work, but rather general findings that are similar throughout all related work.
For this, we focus on Scenario\,\circled{a} (testbed-only) and test whether the performance of out-of-box CC algorithms in our pure-testbed scenario is similar to the findings of related work as presented in \sref{sec:bg}, especially regarding inter- and intra-protocol fairness.

\begin{figure}
\centering
\includegraphics{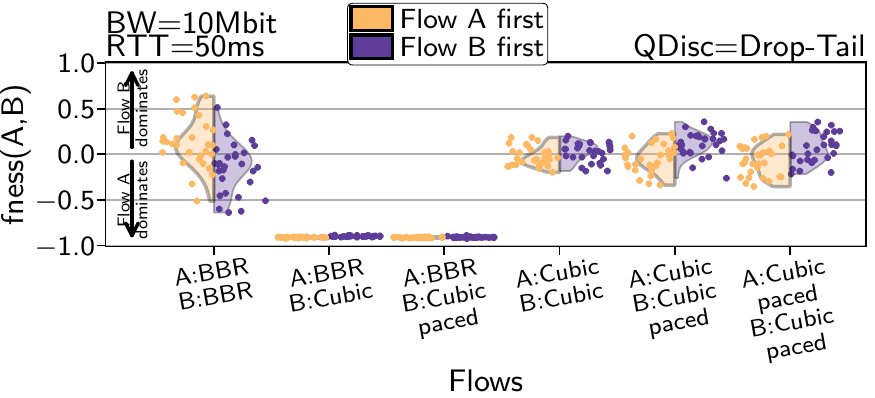}
\caption{Results from Scenario\,\protect\circled{a} for \unit[10]{Mbps}, \unit[50]{ms} and a buffer of 0.5$\times$BDP for different CC algorithm combinations.}
\label{fig:scenario-a-10mbps50ms}
\vspace{-1em}
\end{figure}

\fig{fig:scenario-a-10mbps50ms} visualizes this measure for a configuration with \unit[10]{Mbps} and a min RTT of \unit[50]{ms} for BBR, Cubic, and a Cubic when activating pacing.
We show a scatterplot of all our measured values together with a kernel density estimate to better visualize the location of the majority of our measured data.
For each combination of algorithms, we plot the results when flow A starts first (yellow) side by side with the switched setting when flow B starts first (violet).
For the tests where a CC algorithm performs against itself, switching which flow starts first only mirrors the data at the 0-axis.

Our results show expected values as all algorithms generally show a large degree of fairness to themselves (intra-protocol fairness) with BBR showing a bit of a larger variance compared to the others.
When comparing the \emph{inter}-protocol fairness, we observe that BBR clearly monopolizes the bandwidth regardless of which flow starts first.
This confirms related work on BBR in low-buffer scenarios~\cite{cardwell16:ietf97iccrg:bbr,hock17:icnp:bbrEval,scholz18:bbrDeeperUnderstanding}.
An additional finding is that pacing seems to decrease fairness when competing with both paced and non-paced Cubic flows.

While these experiments validate that our testbed yields meaningful results confirming known findings, we now investigate Scenario\,\circled{b} and Scenario\,\circled{c} to study how CPs, and thus possibly non-standard algorithms from the Internet, perform against our known CC algorithms and against each other.

\section{Congestion Control in the Wild}
\label{sec:results}

We base our evaluation of TCP fairness on actual Internet traffic by six major CPs (Akamai, Amazon, Cloudflare, Edgecast, Fastly, and Google) in two settings: {\em i)} lab vs.\ CP and {\em ii)} CP vs.\ CP in February 2019.
Studying actual Internet traffic is motivated by the observation that CC research often neglects the complex parameterization possibilities.
For example in a previous study~\cite{rueth18:tma}, we found that CDNs use different initial window configurations and some utilize pacing.
To this end, we suspect that not only the initial windows might be different, thus choose two URLs for Akamai (named AkamaiA (then using IW32) and AkamaiE (then using IW16)) mapping to these different settings.
Furthermore, Cloudflare and Google have both publicly announced to utilize BBR.
Thus, we opt to observe the performance of actual Internet traffic originating from these six different CPs when competing against our testbed flows in Scenario\,\circled{b} and against themselves in Scenario\,\circled{c}.

We obtain URLs generating large responses (the smallest being \unit[25]{MB}) served by each CP by analyzing the HTTPArchive.
Since the responses can still be too small to cover our \unit[45]{s} measurement period, we make use of HTTP/2 multiplexing, \ie{} we request the same resource multiple times (in parallel) over the same connection enabling us to prolong the transmission by a multiple of the original file size.
This functionality is already provided by the h2load tool in nghttp2\footnote{\url{https://github.com/nghttp2/nghttp2}}.

\subsection{Lab Traffic vs. Content Provider Traffic}
We start by investigating Scenario\,\circled{b} where traffic from our testbed machines, \ie{} BBR and Cubic flows, competes with Internet traffic and hence the algorithms employed by our CPs.

\begin{figure}[t]
\centering
\includegraphics{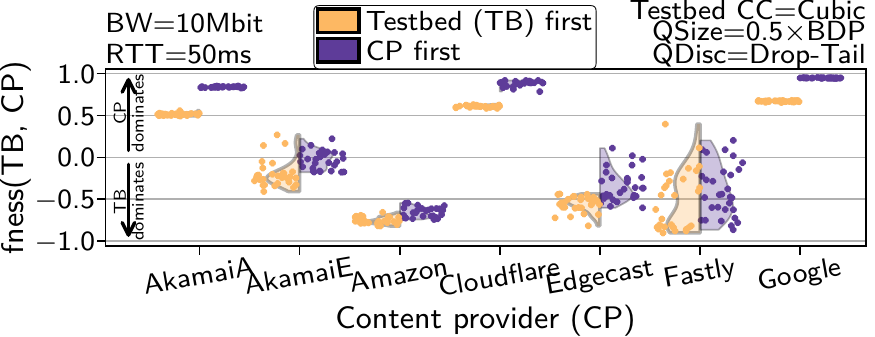}
\includegraphics{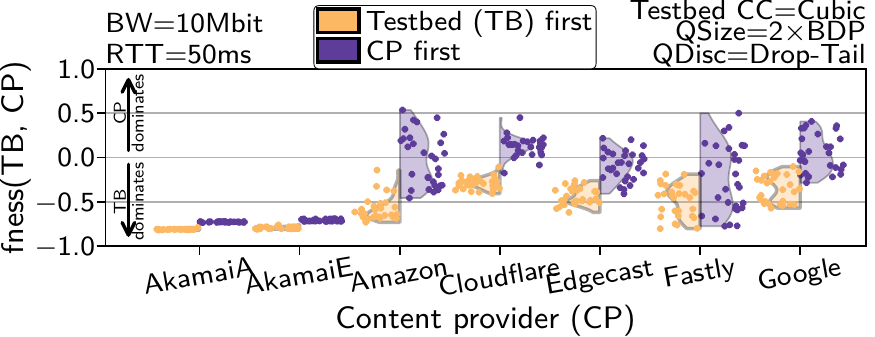}
\caption{A Cubic flow originating from our testbed competes with the content providers for traffic using a small buffer (top) and a large buffer (bottom).}
\label{fig:lvi:cubic1050}
\vspace{-1em}
\end{figure}

\afblock{Cubic Small Buffer.}
\fig{fig:lvi:cubic1050} shows the results for \unit[10]{Mbps} and a min RTT of \unit[50]{ms} when using a Cubic flow.
We again use our fairness measure to plot the measurement results; here, results $< 0$ indicate a dominance for our testbed flow while results $ >0$ favor the CP.
As observed in the top plot for measurements with a small buffer of 0.5$\times$BDP, Cloudflare and Google clearly dominate the traffic in all instances giving little bandwidth to our Cubic flow (unfair setting).
Apart from these two, Amazon and Edgecast struggle against our Cubic flow even when their flow starts first (unfairness by our testbed flow).
In contrast, Fastly---at least when having a headstart---is able to achieve rough fairness.
The two Akamai flows offer a different behavior with AkamaiE showing the highest degree of fairness while AkamaiA is similar to Cloudflare and Google in that it completely dominates our testbed Cubic flow.
This observed difference in behavior of the two Akamai flows supports our initial guess that Akamai uses different configuration parameters.

\afblock{Cubic Large Buffer.}
When looking at the large buffer setting in the plot below we observe a different picture.
Now, the Cloudflare and Google flows do not dominate anymore, the fairness heavily depends on which flow was initiated first.
Similarly for Amazon, Edgecast, and Fastly, when the testbed initiates the first flow, they struggle to gain enough bandwidth.
In contrast, when the CP initiates the first flow, a generally fairer distribution is achieved.
For Amazon and Fastly, we observe a bi-modal distribution of the traffic shares, one that more closely matches the testbed-first case and another that tends to favor the CP.
What is very interesting to see is that both Akamai flows are completely dominated by the testbed Cubic flow, no matter which flow is started first.

\begin{figure}[t]
\centering
\includegraphics{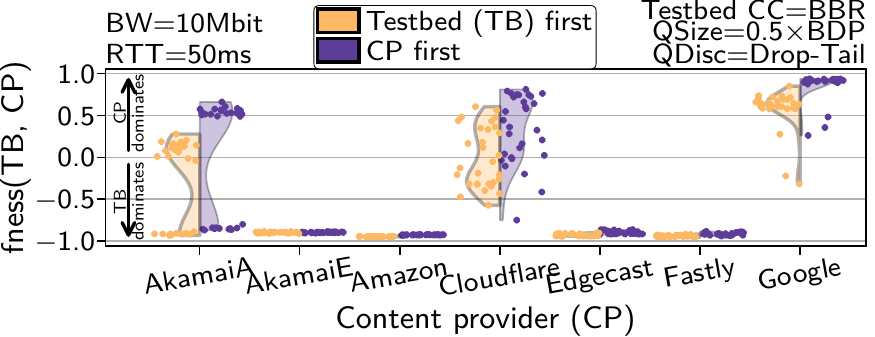}
\includegraphics{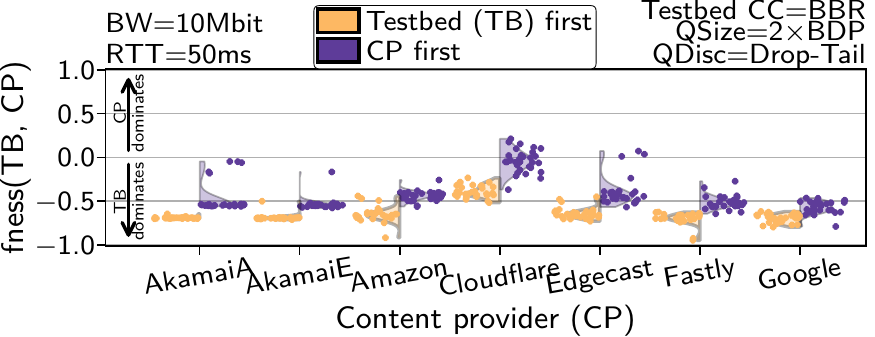}
\caption{A BBR flow originating from our testbed competes with the content providers for traffic using a small buffer (top) and a large buffer (bottom).}
\label{fig:lvi:bbr1050}
\end{figure}

\afblock{BBR Small Buffer.}
Things start to significantly differ when we configure our testbed flow to utilize BBR as shown in \fig{fig:lvi:bbr1050}.
As can be seen in the upper plot, showing the fairness under a small buffer setting, the testbed BBR flow dominates the flows of Amazon, Edgecast, and Fastly.
The same is true for AkamaiE while AkamaiA shows an interesting behavior, as part of the experimental iterations also show the clear dominance of the testbed flow while about half of the iterations either show a very fair result (when the testbed is started first) or a dominance of the Akamai flow (when Akamai is started first).
The observed characteristics hereby seem to be stable in that the behavior seems to switch between two distinct states.
Cloudflare shows a wide range of observed fairness ratios from dominating the testbed flow to the opposite.
For Google, however, our testbed flow always clearly loses to the CP.

\afblock{BBR Large Buffer.}
In the large buffer scenario, the flows of most CPs show very similar behavior in that the testbed flow dominates the competition.
The effect is hereby most visible when the testbed flow starts first while it is slightly ameliorated when the CP is the first flow.
The degree of unfairness is hereby similar in most cases with Cloudflare being the only real exception as it achieves a balanced fairness level when the CP flow is started first.

\begin{table}[t]
\centering
\footnotesize
\setlength{\tabcolsep}{2pt}
\begin{tabular}{@{}l|ll|ll|ll|ll@{}}
\toprule
 & \multicolumn{2}{c|}{BBR@2BDP} & \multicolumn{2}{c|}{BBR@.5BDP} & \multicolumn{2}{c|}{Cubic@2BDP} & \multicolumn{2}{c}{Cubic@.5BDP} \\
 & \multicolumn{1}{c}{QSize} & \multicolumn{1}{c|}{Retrans} & \multicolumn{1}{c}{QSize} & \multicolumn{1}{c|}{Retrans} & \multicolumn{1}{c}{QSize} & \multicolumn{1}{c|}{Retrans} & \multicolumn{1}{c}{Qsize} & \multicolumn{1}{c}{Retrans} \\ 
 \midrule
AkamaiA & 43 & 175 & 14 & 1244 & 60 & 22 & 16 & 80 \\
AkamaiE & 42 & 210 & 13 & 967 & 59 & 24 & 12 & 70 \\
Amazon & 53 & 468 & 12 & 836 & 66 & 30 & 12 & 31\\
Cloudflare & 40 & 22 & 13 & 1319 & 57 & 40 & 12 & 166  \\
Edgecast & 53 & 377 & 12 & 810 & 64 & 33 & 12 & 41 \\
Fastly & 53 & 442 & 11 & 741 & 65 & 31 & 12 & 41 \\
Google & 40 & 215 & 15 & 760 & 55 & 50 & 14 & 184  \\ \bottomrule
\end{tabular}
\caption{Average queue size (QSize) and retransmissions (Retrans) of the testbed originating flows for the \unit[10]{Mbps}, \unit[50]{ms} scenario with the testbed flow starting first.}
\label{tab:lvi:queueStats}
\vspace{-2em}
\end{table}

\begin{figure*}
\includegraphics{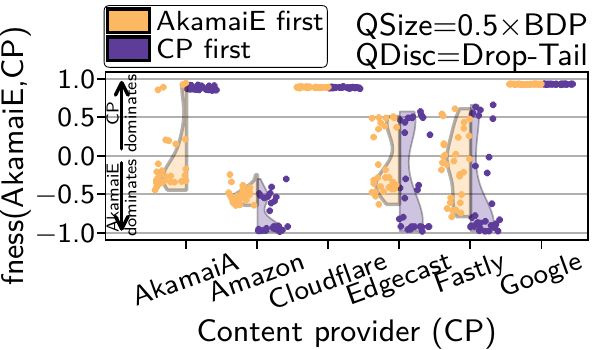}
\includegraphics{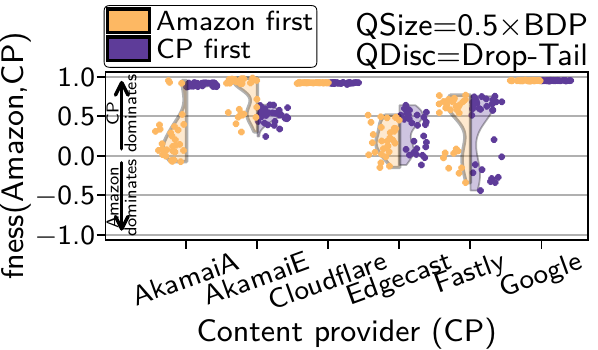}
\includegraphics{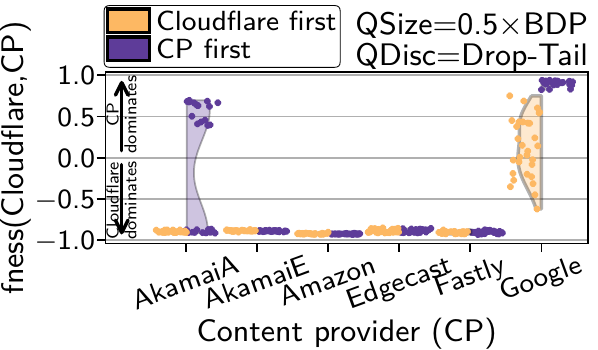}
\includegraphics{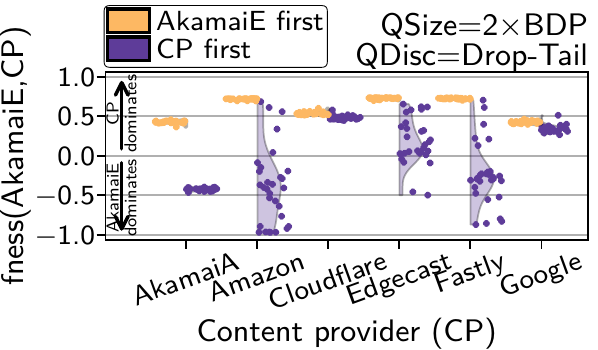}
\includegraphics{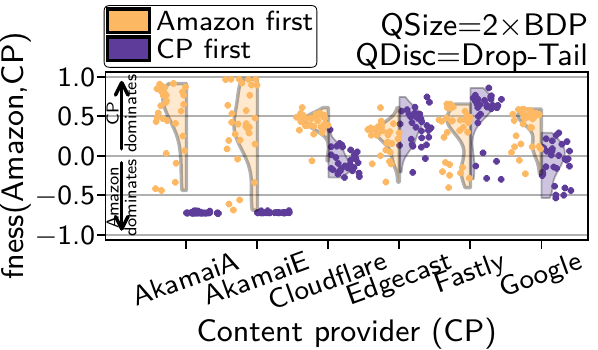}
\includegraphics{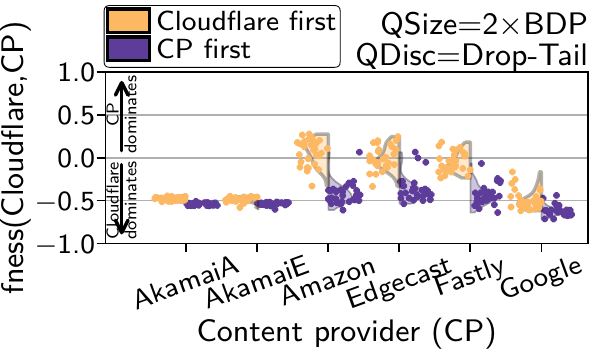}
\caption{AkamaiE, Amazon, and Cloudflare competing against the other content providers in the \unit[10]{Mbps}, \unit[50]{ms} setting. 
Amazon performs similar to Edgecast and Fastly, Cloudflare is similar to Google.
Top row shows 0.5$\times$BDP, bottom row 2$\times$BDP.}
\label{fig:ivi:droptail1050}
\end{figure*}

\afblock{Retransmissions.}
In addition to only looking at the resulting fairness, we also consider key characteristics of the bottleneck buffer and the participating end-hosts.
In this case, we observe the queue size of the bottleneck buffer, which we measure using a simple eBPF program on the bottleneck machine, and the amount of retransmissions of the testbed flow, which we also measure using an eBPF program on the corresponding testbed machine, \ie{} \emph{Testbed1} in \fig{fig:testbed_overview_complete}.
We present these characteristics in Table~\ref{tab:lvi:queueStats}.
What can be seen is that for a testbed Cubic flow, Cloudflare and Google cause significantly higher retransmission counts than the other CPs.
What is very interesting is that Cloudflare induces very few retransmissions for the BBR testbed flow in the large buffer scenario but the most retransmissions in the small buffer scenario.

\afblock{Higher RTT and Higher Bandwidth.}
When we investigate our other settings with larger RTTs we observe no qualitative difference in fairness for all but AkamaiA. 
AkamaiA's bimodal fairness distribution in the 50\% BDP setting shifts towards the testbed dominating all measurements.
When increasing the bandwidth, testbed BBR flows still dominate but the fairness focusses for Cloudflare and Google, especially for smaller buffer sizes; the larger buffer generally leads to a larger distribution of the fairness.
Especially, Amazon, Edgecast, and Fastly can claim slightly more bandwidth on average.
Looking at changes for testbed Cubic flows, we observe no significant difference when competing against Cloudflare and Google. 
For the others, we observe a slight trend towards more bandwidth for the CPs.
Again, AkamaiA stands out in the small queue setting and behaves like AkamaiE when increasing the bandwidth.
We validated AkamaiA's behavior over several days (repeating the same 30 measurements for the different settings) and were able to consistently observe the same changes.

\takeaway{As indicated by our results, fairness largely depends on the available buffer size. 
Generally, it seems that the CC algorithms employed by the CPs are achieving better fairness with off-the-shelf algorithms when more buffer size is available.
However, large buffers can cause jitter and generally inflate the latency.
Yet in small buffer settings, BBR currently claims nearly all bandwidth and shows a large variability in fairness and performance when competing with other BBR flows causing unpredictable performance.}

While it might seem advantageous at first glance that algorithms like BBR claim more bandwidth, it could actually be bad for CPs.
In the web, CPs often compete with 3rd party resources loaded on the same website.
When the CP claims all bandwidth, it may negatively affect the web page loading behavior since they could cause reduced performance for the competing flows of the other resources.
Thus, CPs should interact fair with their competitors which is the focus of the next part of our study, \ie{} how two CP flows interact.

\subsection{Content Provider vs. Content Provider}

For investigating the interaction between the different CPs, we now deploy Scenario\,\circled{c} where both flows are requested from the CPs.
The rest of the testbed configurations remain unchanged.
\fig{fig:ivi:droptail1050} shows the results for AkamaiE, Amazon, and Cloudflare flows competing against the other CPs in a scenario with \unit[10]{Mbps}, a min RTT of \unit[50]{ms} and a small (top) or large (bottom) buffer size.
Due to the similarity of the results, Amazon also serves as a representative for Edgecast and Fastly, while Cloudflare also represents Google.
Once more using our fairness measure, results $< 0$ indicate a dominance of the explicitly mentioned CP while results $ >0$ favor the competing CP mentioned on the x-axis.

\begin{figure*}
\includegraphics{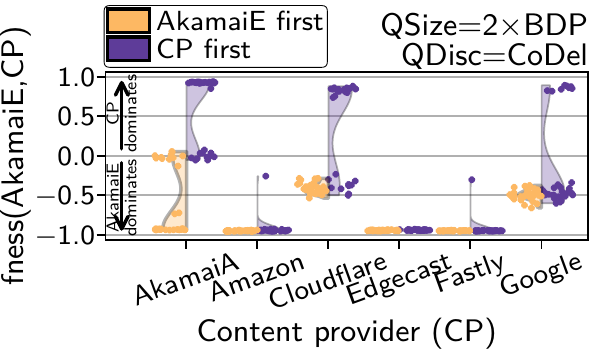}
\includegraphics{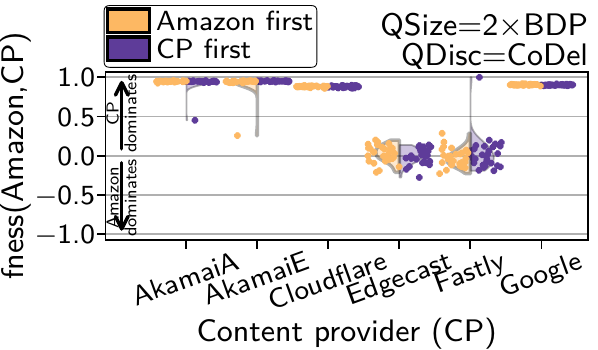}
\includegraphics{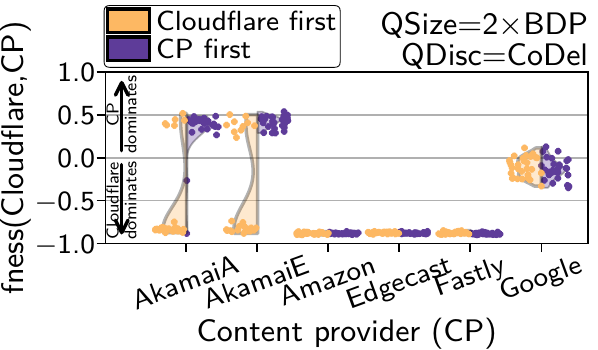}
\includegraphics{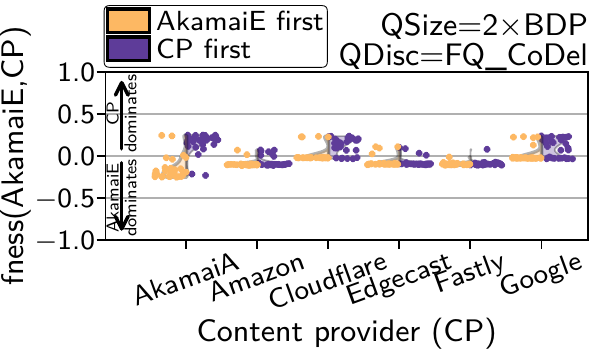}
\includegraphics{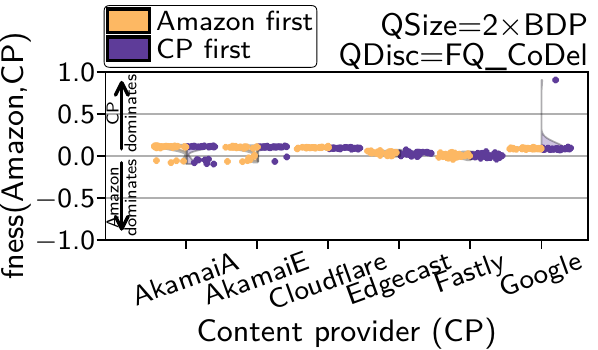}
\includegraphics{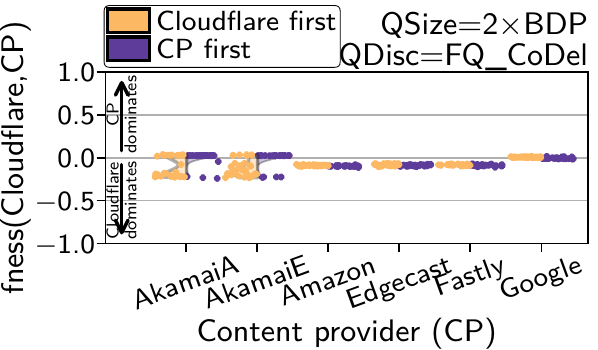}
\caption{AkamaiE, Amazon, and Cloudflare competing against the other content providers in the \unit[10]{Mbps}, \unit[50]{ms}, 2$\times$ BDP setting using CoDel (top) and FQ\_CoDel (bottom).
Amazon performs similar to Edgecast and Fastly, Cloudlfare is similar to Google.}
\label{fig:ivi:codel1050}
\end{figure*}

\afblock{Small Buffers.}
Starting with the upper row, \ie{} with the small buffer scenario, we rarely observe cases where the CPs achieve a good level of fairness.
Especially Cloudflare (right) seems to dominate most of the other CPs with the only exception being when it is forced to compete with Google and AkamaiA.
In the former case, Google generally dominates Cloudflare when it starts first while we observe large range fairness results when Cloudflare is the first flow.
The most interesting observation can be made for AkamaiA as the bi-modal behavior observed before is again visible when it is started first and forced to compete with Cloudflare.
Here, roughly half of the results indicate significant domination by AkamaiA.

For the scenarios where we focus on AkamaiE (left) and Amazon (middle), it is obvious that they are massively dominated by Cloudflare and Google.
The same holds when they compete against a first flow originating from AkamaiA, while the behavior is much fairer when the AkamaiA flow is started second.
When interacting, Amazon, Edgecast, and Fastly show a rather high degree of fairness.
When AkamaiE competes against Edgecast and Fastly, a large range of observed fairness values can be seen, ranging from medium dominance of Edgecast and Fastly to total domination of AkamaiE.
The latter, \ie{} total domination of AkamaiE, is above all visible when competing against Amazon. 

\afblock{Large Buffers.}
Things again change when we focus on the bottom row using a larger buffer.
Regarding Cloudflare (right), we observe that the strict dominance is less profound than in the small buffer scenario yet still favoring it.
When the Cloudflare flow starts first, there is a higher degree of fairness when competing against Amazon, Edgecast, and Fastly, yet they struggle when Cloudflare's flow starts first.
However, Cloudflare does not seem to cooperate well with the two Akamai flows or Google, as it dominates them in all these scenarios.

A generally decent amount of fairness can be observed in several scenarios involving Amazon (middle).
Especially when competing against Cloudflare, Edgecast, and Google, high fairness levels are achieved.
In contrast to that, we can again see very poor fairness for the Akamai flows if they are started first, while we observe a large range of values when they compete against an Amazon flow starting first. 
When the AkamaiE flow is the first contester (left), it is dominated by the other CPs while it seems to be able to better claim bandwidth when it enters as the second flow.

\takeaway{As observed earlier, larger buffers seem to enable a better level of fairness even though they are still far from being equal in most cases.
This is especially true for Cloudflare/Google which dominate most of the other CPs in the small buffer scenario while there are reasonable fairness values for most CPs in the large buffer setting.
}

Even though a higher level of fairness can be noted for the large buffer, it comes with the problem of larger queue sizes and hence also with increased delays.
Ideally, we would have a scenario with a smaller queue but still the high level of fairness.
As AQMs like CoDel are designed to keep the delay (and hence the queue) small, we are interested in whether they can help to achieve the desired combination of small delays and high level of fairness.
This is why we investigate the effect of AQMs on the whole situation in the following section.

\subsection{Can CoDel Improve Fairness?} %
\label{sub:can_codel_improve_fairness_}

AQMs inherently change the behavior of a queue which is why they have a significant impact on the overall performance.
Generally, they have two possible forms of feedback to which flows might respond: i) dropping packets and ii) using ECN.
In our work, we only consider the first case of feedback, \ie{} packet drops because it requires no end-to-end support.
For this, we repeat the experiments from before but activate CoDel and its flow-queuing variant FQ\_CoDel on the intermediate bottleneck machine.
In the following, we further concentrate on the case with a queue size of 2$\times$BDP because CoDel's effect on small queues is likely to be diminishing.
Hence, \fig{fig:ivi:codel1050} only shows the results for a queue size of 2$\times$BDP in the otherwise unchanged scenarios previously used in \fig{fig:ivi:droptail1050}, \ie{} for \unit[10]{Mbps} and a min RTT of \unit[50]{ms}.
We again choose AkamaiE, Amazon, and Cloudflare as the showcase CPs.

\afblock{CoDel.} 
The main observation that can be made is that CoDel (top) seems to achieve a very high level of fairness when Amazon competes with Edgecast and Fastly and when Cloudflare competes with Google.
Apart from that, there are above all very bad fairness values when other CPs are competing against Amazon or Cloudflare with tendencies looking like the small buffered scenario with a FIFO queue.

For the AkamaiE flow, the application of CoDel comes in hand with a clear domination of Akamai when competing against Amazon, Edgecast, and Fastly.
What is more, when it is started first, Akamai also dominates Cloudflare and Google, while there are again two regions of values when Cloudflare and Google are started first; one where Akamai dominates and one where the other two dominate.
Finally, when looking at the performance against AkamaiA, the two-regional effect is again visible and now for both cases.
AkamaiE starting first is hereby characterized by a dominance of AkamaiE half of the time and a fair behavior the other half, while this is the exact opposite if AkamaiA is started first.

\afblock{FQ\_CoDel.} 
Shifting towards the flow-queuing variant (bottom), which is designed to produce a fair queuing, we observe a tremendous increase in fairness.
Now, throughout all measurements, fairness is close to the equilibrium and we only observe slight variations.
When looking at AkamaiE we see the largest variation relative to the others with AkamaiE slightly dominating most of the others.
Looking at Amazon, we see a slight advantage that diminishes for Edgecast and Fastly.
In the Cloudflare case, Amazon, Edgecast and Fastly get slightly less bandwidth while Google is very fair and the Akamai's again showing a slight bimodal pattern.

\takeaway{Combining the findings of this section with our previous observations that Amazon, Edgecast, and Fastly use a similar algorithm and that Cloudflare and Google use BBR, it can be said that CoDel above all seems to improve the intra-protocol fairness in large buffers. 
This is bad news for the heterogeneous Internet, as scenarios with different algorithms suffer from severe unfairness.
Luckily, the flow-queuing variant enables a large degree of fairness even in heterogeneous settings.
Thus, it seems to again stand that the technologies to enable a fair and performant Internet are available and only need to be deployed at the bottlenecks.
}

\section{Conclusion}
\label{sec:conclusion}
In this work, we empirically investigated the fairness of content providers in the Internet.
With the help of our testbed, we are able to investigate actual Internet traffic subject to RTT-fairness when competing under lab-controlled properties of a bottleneck.
Generally, we find there is only limited fairness in the Internet today.
Some content providers interact well with each other, while others do not which is likely reflected in their choice of congestion control algorithm.
We find that the bottleneck buffer size significantly impacts the fairness enabling it to invert observations when going from small to large.
This demands research to shine a light on actual configurations of bottleneck buffer sizes in the Internet to then investigate, \eg{} the impact on web performance when content is served from a diverse set of content providers.
Still, there is a silver lining: State-of-the-art AQMs such as FQ\_CoDel put the fairness control back into the network operator's and possibly the end-user's hand, however, require deployment on millions of devices.

\afblock{Acknowledgment.}
Funded by the Excellence Initiative of the German federal and state governments, as well as by the German Research Foundation (DFG) as part of project B1 within the Collaborative Research Center (CRC) 1053 -- MAKI.

\bibliographystyle{IEEEtran}
\bibliography{paper}

\end{document}